\documentclass[usenatbib]{mn2e}
\usepackage[dvips]{graphicx}
\usepackage{amssymb}

\title[Long term variability in LMXB 4U1636-536]{Correlated Optical/X-ray Long-term Variability in LMXB 4U1636-536}

\author[I. C. Shih]{I. C. Shih$^{1,2}$\thanks{E-mail: ichsih@phys.nthu.edu.tw}, P.A. Charles$^{3,4}$, R. Cornelisse$^{5}$ \\
$^{1}$Department of Physics \& Astronomy, Michigan State University, East Lansing MI 48824, U.S.A.\\
$^{2}$Institute of Astronomy, National Tsing Hua University, Hsinchu, Taiwan\\
$^{3}$South Africa Astronomical Observatory, Cape Town, South Africa\\
$^{4}$School of Physics \& Astronomy, University of Southampton, Southampton, UK\\
$^{5}$Instituto de Astrofisica de Canarias, Tenerife, Spain}

\begin{document}

\date{}

\pagerange{\pageref{firstpage}--\pageref{lastpage}} \pubyear{}

\maketitle

\label{firstpage}

\begin{abstract}
We have conducted a 3-month program of simultaneous optical, soft and hard X-ray monitoring of the LMXB 4U1636-536/V801 Ara using the SMARTS 1.3m telescope and archival RXTE/ASM and Swift/XRT data. 4U1636-536 has been exhibiting a large amplitude, quasi-periodic variability since 2002 when its X-ray flux dramatically declined by roughly an order of magnitude. We confirmed that the anti-correlation between soft (2-12 keV) and hard ($>$ 20 keV) X-rays, first investigated by \citet{icshih2005}, is not an isolated event but a fundamental characteristic of this source's variability properties. However, the variability itself is neither strictly stable nor changing on an even longer characteristic timescale. We also demonstrate that the optical counterpart varies on the same timescale, and is correlated with the soft, and {\it not} the hard, X-rays. This clearly shows that X-ray reprocessing in LMXB discs is mainly driven by soft X-rays. The X-ray spectra in different epochs of the variability revealed a change of spectral characteristics which resemble the state change of black hole X-ray binaries. All the evidence suggests that 4U1636-536 is frequently ($\sim$monthly) undergoing X-ray state transitions, a characteristic feature of X-ray novae with their wide range of luminosities associated with outburst events. In its current behavioural mode, this makes 4U1636-536 an ideal target for investigating the details of state changes in luminous X-ray binaries.
\end{abstract}


\section{Introduction}
The low mass X-ray binary (LMXB) 4U1636-536 is a well-known type-I X-ray burster, which has been extensively observed over the past four decades and is considered well-established as a persistent X-ray source \citep[see references in][]{Liu2007}.  Its optical counterpart (V801 Ara) revealed its 3.8h orbital period \citep{Smale1988} through photometric variations, but the period has also been detected spectroscopically \citep{Casares2006}.  However, during 2001, the {\it All Sky Monitor} onboard the {\it Rossi X-ray Timing Explorer} (RXTE/ASM) recorded an obvious, gradual decline of its X-ray flux level. Within a year, it began to exhibit a substantial quasi-periodic, long-term variability of $\sim$46d \citep{icshih2005}. More remarkably, Shih et al. also reported that this variation was present at both soft (2-12 keV, RXTE/ASM) and hard (20-100 keV, INTEGRAL/IBIS) X-ray energies, but that these were {\it anti-correlated}. Such a pronounced energy-dependent phenomenon led to the suggestion of an accretion disc instability as the most likely explanation, as the result of a decrease in the mass transfer rate.\\

\cite{Belloni2007} provided further evidence regarding the nature of the long-term variability of 4U1636-536 based on a $\sim$1.5 yr RXTE/PCA monitoring campaign from March 2005 to September 2006. The lightcurves display long-term variability of 30-40d which is consistent with the temporal and energy changes that occur between two states of atoll sources: the {\it island} (hard) and {\it banana} (soft) states \citep{Hasinger1989}. From recently published Swift/BAT hard X-ray (20-50 keV) light curves that began in 2005, \citet{Farrell2009} have confirmed the anti-correlated relationship between the soft and hard X-rays in 4U 1636-536, providing further support for the suggestion of a state change/accretion instability within the source.\\

According to the standard model for LMXBs,  their optical emission is dominated by X-ray reprocessing in the accretion disc around the compact object \citep{vanParadijs1994}. Thus, the varying X-ray flux of 4U1636-536 should also influence the apparent brightness of its optical counterpart. With this in mind, we began a $\sim$100d optical daily monitoring program which was expected to cover at least 2 cycles of the previously observed long term X-ray variability in 4U1636-536.\\

In this paper, we report the results of our optical observations, and compare them with the simultaneous X-ray data available from the RXTE/ASM and Swift/BAT archives. These X-ray monitoring data also allowed us to investigate the long-term variability properties of 4U 1636-536 as a function of X-ray energy. An X-ray spectroscopic analysis of RXTE/PCA and HEXTE data was carried out to investigate the evolution of its spectral properties as a function of the long-term variability. We also discuss possible physical mechanisms that might drive such phenomena.\\

\section{Observations}
\subsection{SMARTS optical campaign}
We performed a 3 months optical monitoring of 4U1636-536 using the NOAO SMARTS 1.3m telescope with the ANDICAM on Cerro Tololo, Chile between 01 April and 30 June 2008. The CCD for the ANDICAM is a Fairchild 447 2048x2048 CCD with 15-micron pixels. Each night, a pair of exposures of 2 minutes each through both B and V filters was taken. In total, we obtained 67 nights of good data during the campaign. Data reductions, such as  bias subtraction and flat-field correction, were provided by the Yale SMARTS team\footnote{http://www.astro.yale.edu/smarts/ANDICAM/}.\\ 

We utilised the aperture photometry package {\tt IRAF/DAOPHOT} to extract photometric information from the data. A number of bright comparison stars (listed in the NOMAD catalog) on the same CCD frame were selected to reconstruct the variability of the source, but absolute calibration of the field was not performed in this work. From the scatter of the differential magnitudes of companion stars, the accuracy of the differential magnitude measurement of 4U 1636-536 is typically $\sim$0.006 mag.\\

\subsection{X-ray monitoring data}
Since the 1960s 4U1636-536 has been frequently observed by many X-ray missions. The All-Sky Monitor (ASM, 2-12 keV) onboard the long-lived Rossi X-ray Timing Explorer (RXTE) has been providing the longest continuous coverage of all luminous X-ray sources, stretching from 1996 to the present day \citep[see][]{Levine1996}.  Additionally, the Hard X-ray Transient Monitor program, using the Burst Alert Telescope (BAT, $\sim$15-50 keV) onboard Swift, has been operating since 2005 \citep{Barthelmy2005}. We retrieved both datasets up to 09 February 2009 from the HEASARC archive\footnote{http://heasarc.gsfc.nasa.gov/docs/archive.html} and the website of the Swift/BAT Hard X-ray Transient Monitor\footnote{http://heasarc.gsfc.nasa.gov/docs/swift/results/transients/}, respectively.\\

\begin{figure*}
\includegraphics[width=170mm]{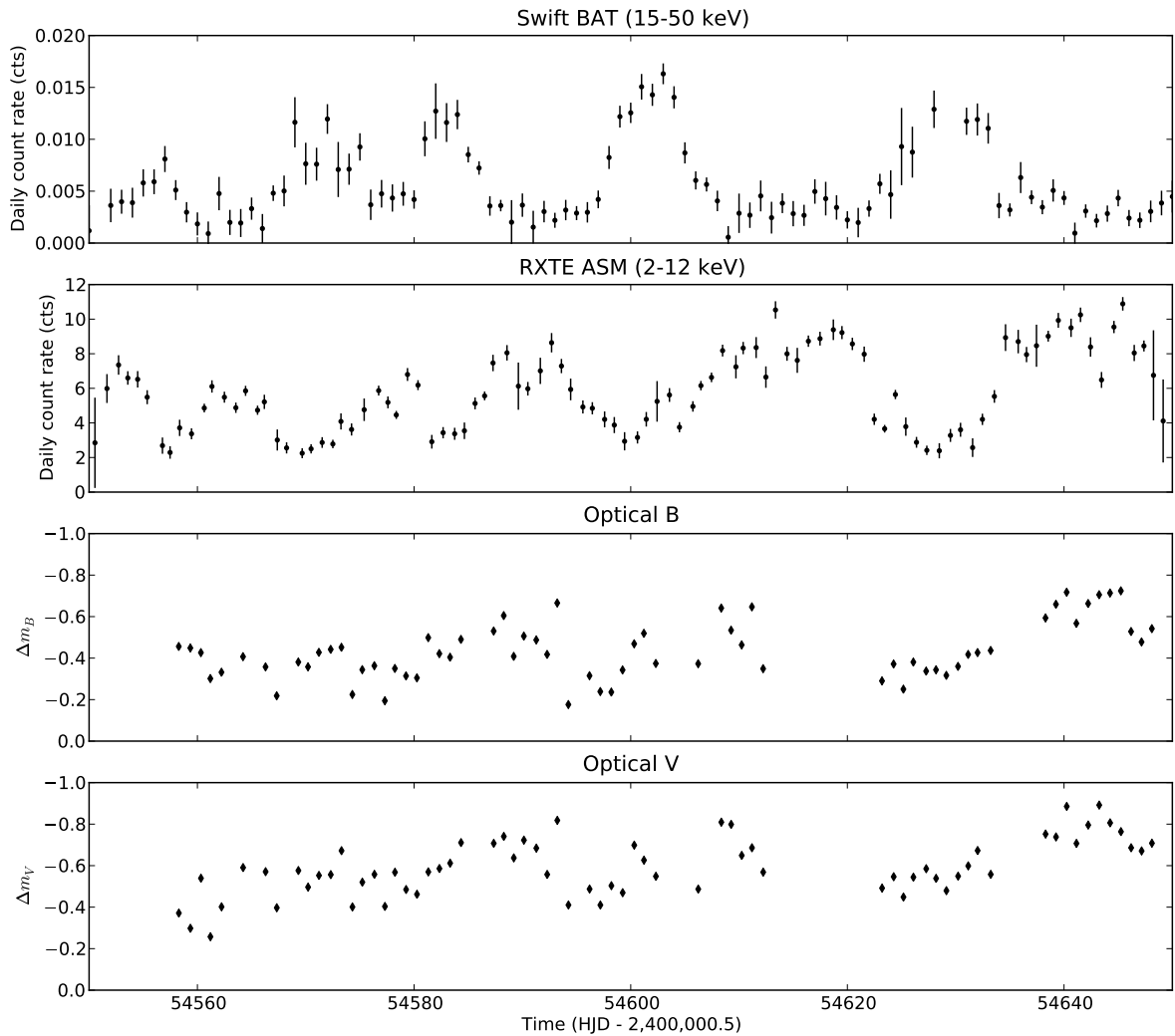}
\caption{From top to bottom, Swift/BAT (hard X-ray), RXTE/ASM (soft X-ray) count rates and SMARTS 1.3m telescope optical (B, V) lightcurves of the LMXB 4U1636-536.  Relativity magnitudes are presented in both optical B and V bands, and the offset is $\sim$17.4.}
\label{multiple lightcurves}
\end{figure*}

\section{Analysis}
\begin{figure}
\includegraphics[width=90mm]{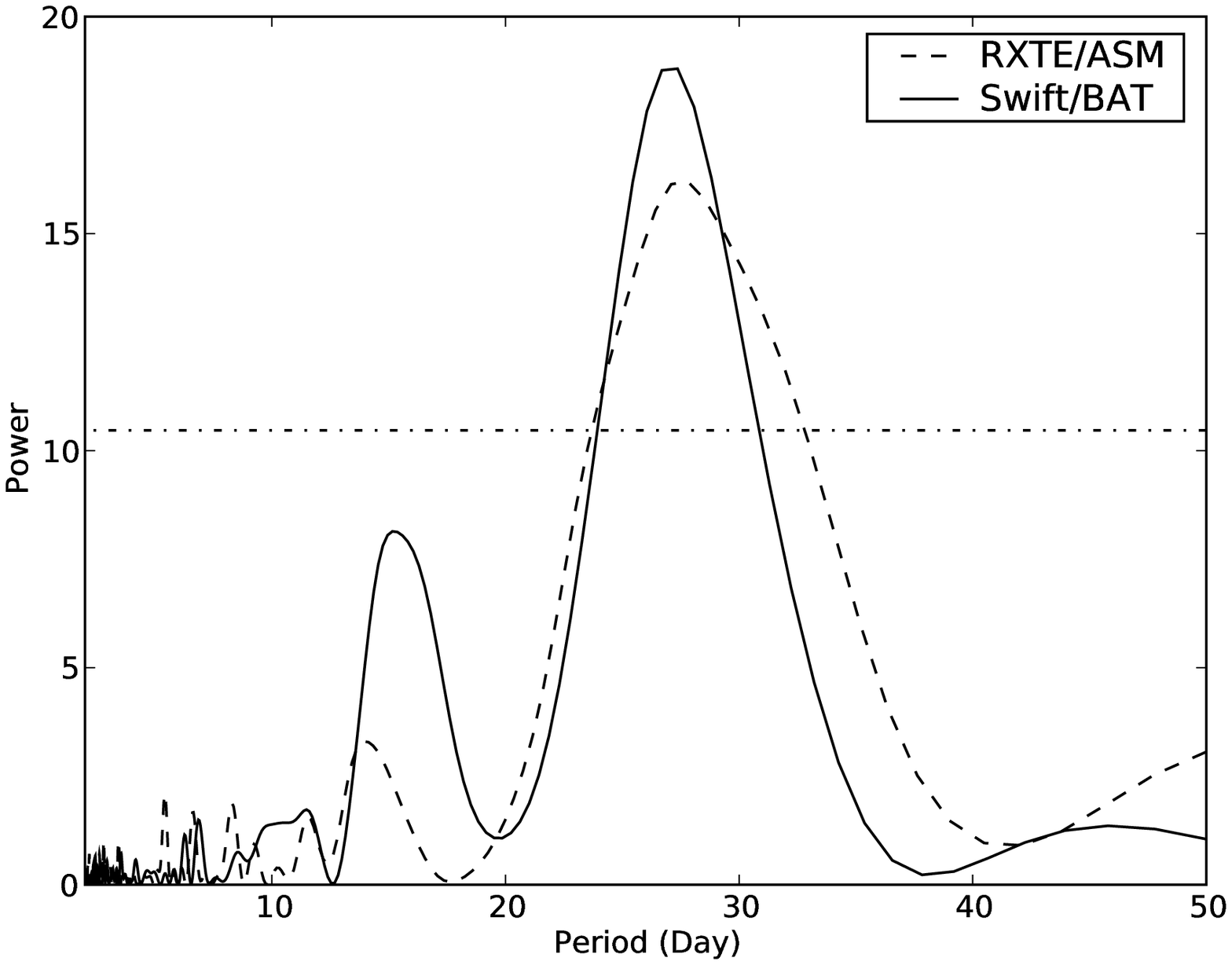}
\includegraphics[width=90mm]{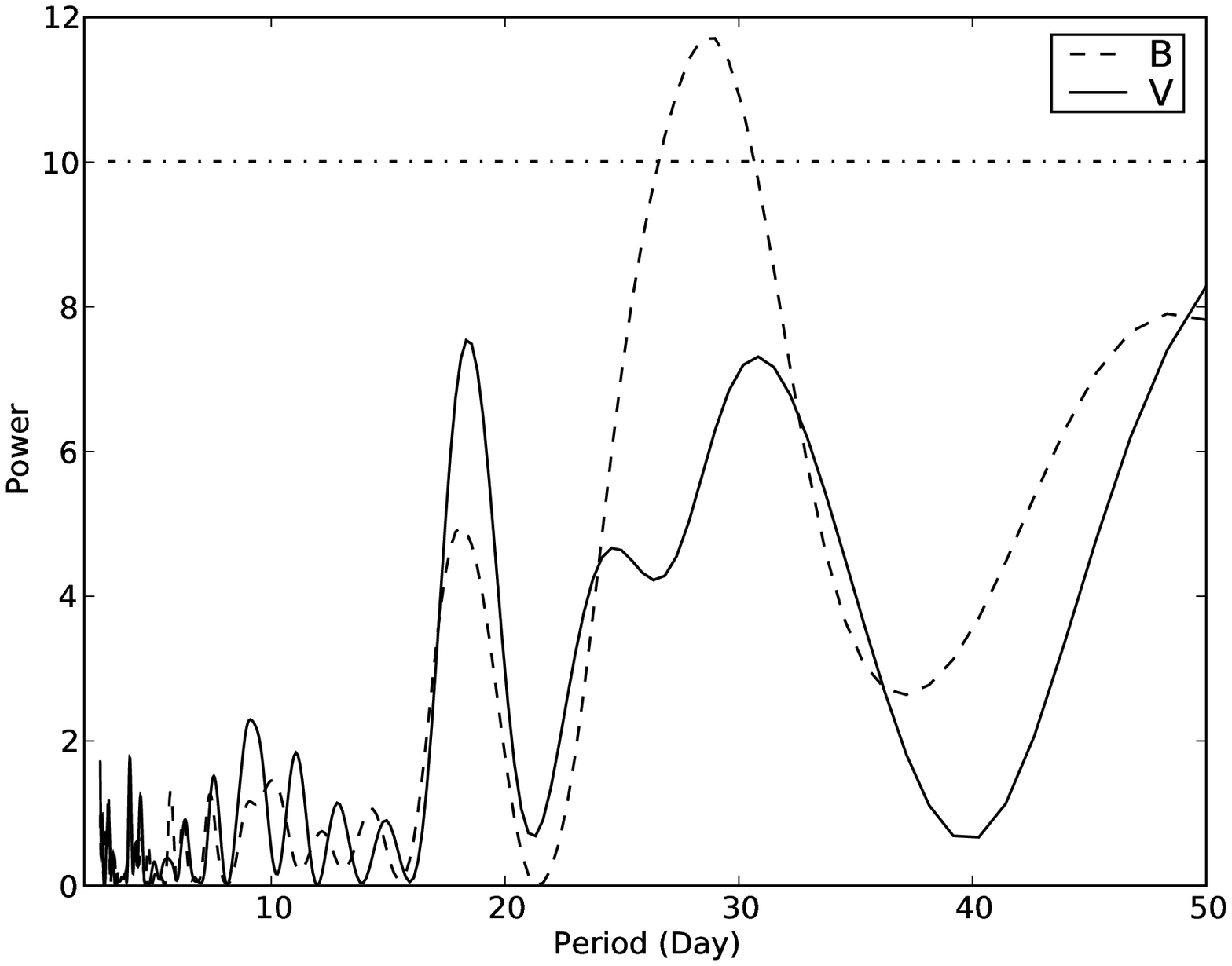}
\caption{Lomb-Scargle periodograms of soft and hard X-rays (upper), and optical B and V band data (lower) for 4U1636-536 in 2008. The dash-dot lines indicate the 99.9$\%$ confidence level (see text).}
\label{LS periodogram}
\end{figure}

Figure \ref{multiple lightcurves} displays the three months of SMARTS monitoring in the optical B and V bands, together with the simultaneous soft (RXTE/ASM) and hard X-ray (Swift/BAT) data. The first feature to note is the presence of a nearly regular variation in all wavelengths, which is even more obvious in the X-ray data. In figure \ref{LS periodogram} the Lomb-Scargle periodogram reveals a significant, consistent period of $\sim$30 d at all wavelengths except the optical V band. We notice that the peak profile is rather wide, so we simulated our data with a 30d sine wave (plus Poisson noise) using a similar count rate as observed for 4U 1636-536. This simulation shows a peak with similar width to that observed, suggesting that it is a result of the short length of the simultaneous dataset (only covering a few cycles). Indeed, the peak width becomes narrower as we increase the duration of data included (RXTE/ASM and Swift/BAT only). For example, within one year surrounding the campaign, the source clearly exhibited a significant long-term variability of $\sim$ 30d in both energy bands.\\

\subsection{Longer term variability}
\begin{figure}
\includegraphics[width=80mm]{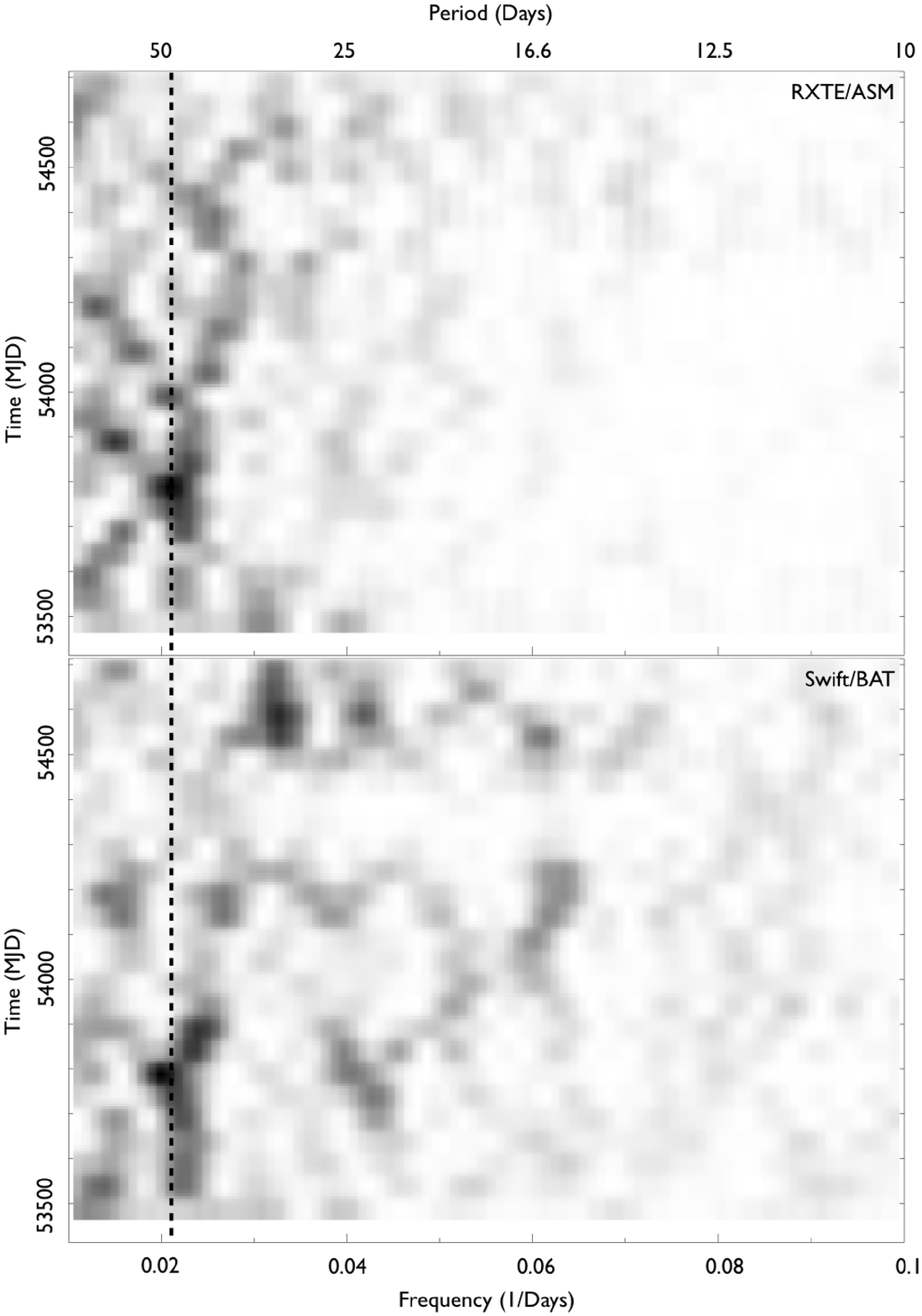}
\caption{The long-term (2005-2009) time-dependence of power spectra of 4U1636-536 from RXTE/ASM (upper) and Swift/BAT (lower) between 2005 and 2009 showing how the quasi-periodic variation evolves.  The $\sim$46d modulation reported by \citet{icshih2005} is marked by the dashed vertical line.}
\label{swift and rxte power spectra}
\end{figure}

The period present in the 2008 campaign was clearly shorter than that exhibited in 2004 \citep[$\sim$46 d,][]{icshih2005}, and was not always consistent with other reports, i.e. 30-40d \citep{Belloni2007} and $\sim$46d \citep{Farrell2009}. Until recently, such variability studies have been rather patchy, and the correlation between soft and hard X-rays was not fully established.  However, with the availability of both RXTE/ASM and Swift/BAT monitoring over extended periods of time, a time-dependent power spectral analysis, similar to that reported by  \citet{Clarkson2003} on the superorbital modulation in SMC X-1, was performed in order to investigate the long-term variability evolution in 4U 1636-536. As the timescale of interest is $\sim$50d, a sliding data window of 200d was used. In order to provide some smoothing of the resulting dynamical power spectrum, each subsequent window moved forward by 50d. The resulting power spectrograms from both datasets are shown in figure \ref{swift and rxte power spectra}.\\

Our analysis suggests that even the dominant periodicity (the darkest region) is transient in nature. The most recent (2009) data show that the variation has evolved to a shorter timescale ($\sim$30d), but our analysis did not reveal any ultra long-term periodic trend, as was seen in e.g. SMC X-1 \citep{Clarkson2003}. Near the three months of SMARTS observations (HJD $\sim$54600), it appears that the $\sim$30 d period was only significant in Swift/BAT data. However, as we have shown in figure \ref{LS periodogram}, the $\sim$30 day period is also significant in RXTE/ASM data.\\

\subsection{Correlations}
To investigate the correlation between different wavebands, we performed a cross-correlation analysis with the three datasets: (i) optical and soft X-rays, (ii) optical to hard X-rays, (iii) soft and hard X-rays. Since the variability is more significant in optical B band, that was used in the analysis (after conversion from magnitude to intensity).\\

\begin{figure}
\includegraphics[width=90mm]{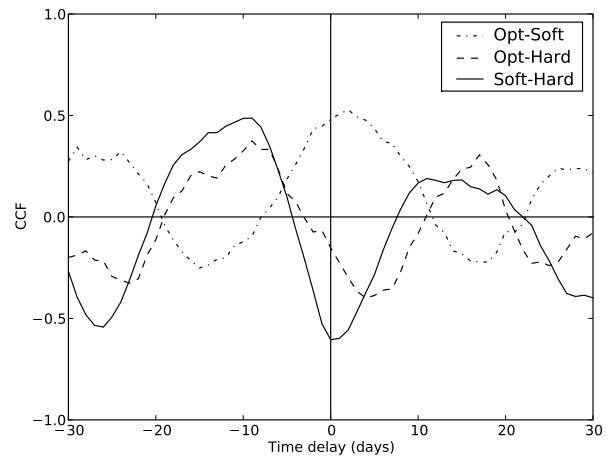}
\caption{Cross-correlation functions of the three 4U1636-536 datasets:  optical to soft X-rays (dot-dashed line); optical to hard X-rays (dashed line); and soft to hard X-rays (solid line).}
\label{cross correlation}
\end{figure}

Our analysis adopted the interpolation method of calculating the cross-correlation function \citep[CCF,][]{White1994}. The light curves were shifted backwards and forwards up to 30 days against each other in steps of 1 day. The results of this analysis are shown in figure \ref{cross correlation}.\\
  
The CCF analysis reveals a clear anti-correlation between the soft and hard X-rays. On the other hand, the optical emission is positively correlated with the soft X-rays. However, we also note that the optical emission seems to lag the soft X-rays by $\sim$ 1-2 days, which is much too long to be the result of X-ray reprocessing/irradiation (time-scale of $\sim$ minutes). The optical emission, like the soft X-rays, is anti-correlated with hard X-rays, and the lag is slightly longer.  However, we caution that the lag appearing in the CCF may not be real, but is instead an artefact related to the gaps in the optical data.  A more regular and longer term optical campaign will be needed to confirm this.\\

\subsection{X-ray spectral analysis}
\begin{figure}
\includegraphics[width=90mm]{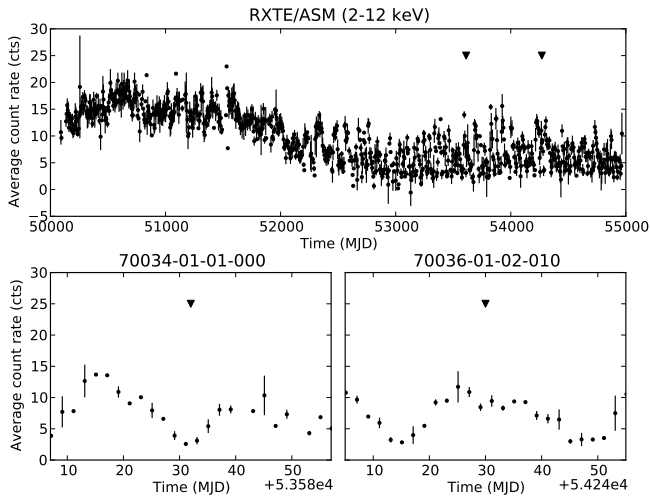}
\caption{(Upper) The complete RXTE/ASM light-curves of 4U1636-536 covering more than 13 years. The data have been rebinned with an interval of 5 days for clarity at this scaling. The black triangles indicate times of RXTE observations selected for detailed spectral analysis (see text). (Below) Details from the original one day average RXTE/ASM data which are in the vicinity of the RXTE/PCA observations.}
\label{RXTE_total}
\end{figure}

In an attempt to understand how the X-ray spectral properties of 4U 1636-536 evolve through the different phases of the  long-term variability, a broad-band spectroscopic analysis was performed.  We used archival RXTE/PCA and HEXTE data because of their wide spectral range between 2 and 250 keV \citep{Jahoda2006, Rothschild1998}. Two observations, which represent two distinct epochs in the RXTE/ASM light-curve (at maximum and minimum) and sufficiently long to provide good signal-to-noise ratio, were selected (see figure \ref{RXTE_total}): 

\begin{itemize}
\item ObsID 70036-01-02-00 (MJD 54270),  at the peak of an oscillation
\item ObsID 70034-01-01-000 (MJD 53612), at the minimum of an oscillation
\end{itemize}

The spectra were derived from PCA Standard-2 and HEXTE Archive Mode following standard procedures: using only PCU 2 (the only PCU that operates in all our datasets) in PCA and Cluster B (which provides the necessary background information) in HEXTE and the latest calibration data \citep{PCARMF}. The combined PCA and HEXTE data were then analysed with {\tt Xspec v.12.x}. Throughout the analysis, the Galactic column density ($N_H$) toward the source was fixed at 3$\times$10$^{21}$ cm$^{-2}$ \citep{Dickey1990}. We note that this value is well within the range determined by the ROSAT Position Sensitive Proportional Counter observations \citep[PSPC,][]{Schulz1999}. To estimate the luminosity and radii (see context below), the distance of 4U 1636-535 is referred to $\sim$ 5.9 kpc \citep[][]{Cornelisse2003}. \\

\begin{table}
\begin{center}\begin{tabular}{lcc} \\
\hline
ObsID & 70036-01-02-00 & 70034-01-01-000 \\
ASM Epoch & high & low \\
\hline
{\tt DiskBB} & & \\ \hline
T$_{in}$ (keV) & 0.65$\pm0.06$ & - \\
nor. & 296.5$\pm^{179.3}_{99.3}$ & - \\ \hline
{\tt Blackbody} & & \\ \hline
kT (keV) & - & 1.14$\pm0.25$ \\
nor. & - & 0.002$\pm0.001$ \\ \hline
{\tt compTT} & & \\ \hline
T$_{0}$ (keV) & 1.05$\pm0.08$ & 1.05 (fixed) \\
kT (keV) & 2.88$\pm^{0.12}_{0.09}$ & 5.15$\pm^{0.90}_{0.85}$ \\
$\tau$ & 5.16$\pm^{0.28}_{0.34}$ & $>$5.13 \\
nor & 0.05$\pm0.01$ & 0.007$\pm^{0.002}_{0.004}$ \\ \hline
{\tt Power-Law} & & \\ \hline
$\Gamma$ & - & 2.12$\pm0.04$ \\
nor. & - & 0.29$\pm^{0.03}_{0.11} $ \\ \hline
{\tt Lorentz} & & \\ \hline
LineE (keV) & - & 6.28$\pm^{0.38}_{0.61}$ \\
Width (keV) & - & 3.74$\pm^{1.89}_{1.85}$ \\
nor. & - & 0.008$\pm^{0.03}_{0.007}$ \\
\hline
{\tt L$_{X}$} (10$^{36}$ erg s$^{-1}$) & & \\ \hline
2.0-80.0 keV & 4.41$\pm$0.02 & 7.50$\pm$0.01 \\
\hline
$\chi_{\nu}^{2}$ & 1.20 & 0.98 \\
\hline
\end{tabular}
\caption{Spectral model fits to two distinct states of 4U1636-536 with the RXTE PCA/HEXTE data. The X-ray luminosity is determined from the unabsorbed flux.}
\label{Spectral Data}
\end{center}
\end{table}

\begin{figure*}
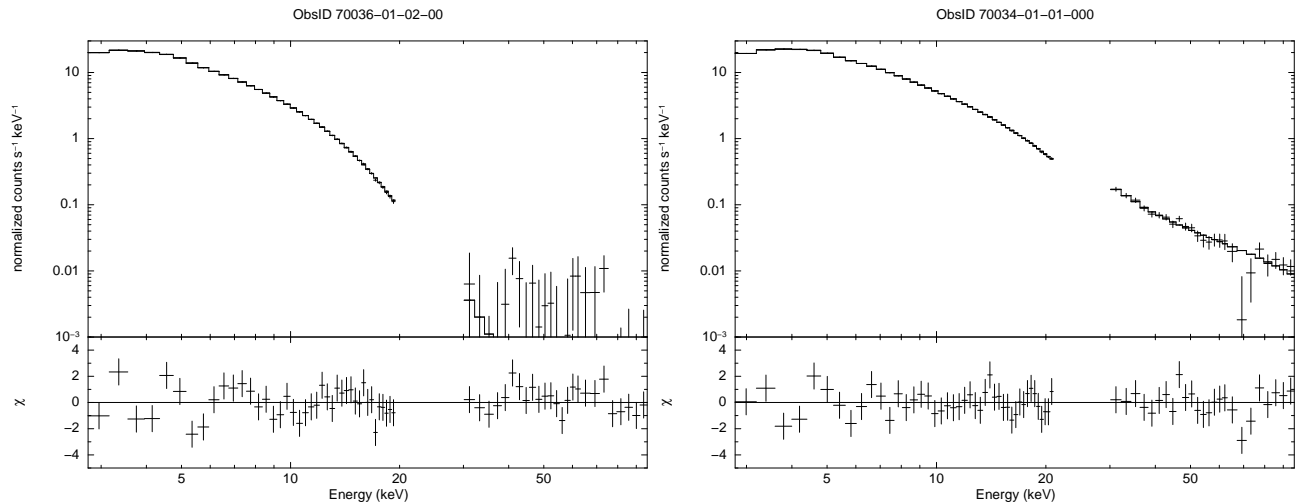

\includegraphics[angle=270, width=85mm]{1636-fit_70036.ps}
\includegraphics[angle=270, width=85mm]{1636-fit_70034.ps}
\caption{Spectra of a bright (left) and low (right) epoch observed by RXTE/PCA and HEXTE, shown together with the total model and residuals with respect to the corresponding best fits.}
\label{spectral fitting result}
\end{figure*}

Figure \ref{spectral fitting result} shows the data and model fits, with table \ref{Spectral Data} summarising the best fit model details. The spectral properties of 70036-01-02-00 (RXTE/ASM high epoch) can be simply described by a model consisting of a multi-temperature accretion disc ({\tt DiskBB}) component plus a thermal Comptonisation ({\tt CompTT}, a spherical geometry was assumed) component.\\

On the other hand, the spectral properties of 70034-01-01-000 (RXTE/ASM low epoch) is more complicated. The same two-component model fit to the data is poor ($\chi_{\nu}^{2} \gg$ 2). Instead, it is more compatible with a single-temperature blackbody component ({\tt Bbody}) plus a thermal Comptonised one (using the same input soft photon temperature T$_{0}$ as in the high-state epoch). The blackbody component likely represents the emission from the surface of the neutron star. Additionally, a power-law component for the hard X-rays $\gtrsim$ 50 keV and a Lorentz profile at $\sim$ 6.3 keV (which may represent a relativistically broadened Fe K$\alpha$ emission line) are required to improve the fitting result. We note that these additional components are quite similar to those observed in the hard state of black hole X-ray binaries, such as GRO J1655-40 \citep[see][]{Remillard2006}.

\section{Discussion}
\subsection{X-ray variability}
\label{X-ray variability}
From the 3 years of simultaneous RXTE/ASM and Swift/BAT archival data on 4U1636-536, it is clear that the variability, on both short and long time-scales, is indeed strongly energy-dependent.  This was first noted in simultaneous RXTE/ASM and INTEGRAL/IBIS data \citep{icshih2005}, and has been a persistent feature of this source since 2002, when RXTE/ASM observed the beginning of a dramatic decline in flux. Additionally, we find that the long-term variability of 4U1636-536 has been constantly evolving over the last decade, although not in a periodic manner.\\

A tilted/warped, precessing accretion disc was suggested as an explanation for such long term variations \citep{icshih2005}, and the system parameters (in particular the low mass-ratio) of 4U 1636-536 do allow for a precessing accretion disc \citep{Casares2006}. However, the presence of an evolving trend in the long-term period, the anti-correlation between soft and hard X-rays, and the distinct X-ray spectral properties at different epochs are all difficult to understand in terms of a precessing disc.\\

The X-ray spectroscopic analysis clearly reveals two different spectral states in the high and low epochs. It suggests that, in the high epoch, 4U1636-536 is like a typical LMXB: the accretion disc is fully developed and extends into the inner boundary region (R$_{in} \sim$ 25 km, assuming face-on orientation, $i$=0$^{\circ}$), which is likely surrounded by a Comptonised region, or corona \citep{Mitsuda1989}. The energy state is softer as the spectrum is dominated by soft X-rays (2-10 keV).\\

In the low epoch, on the other hand, no reliable multi-temperature disc component could be derived suggesting that the accretion disc probably retreated away from the inner region, or was {\it truncated}. Instead, we have determined a soft X-ray emitting region of radius $\sim$ 5 km, which most likely corresponds to the neutron stellar surface. We also note that the spectral data in the low epoch contains a high energy component ($\gtrsim$50 keV) which is very similar to that in a ``peculiar'' INTEGRAL observation of 4U1636-536 \citep{Fiocchi2006}, as well as to those of black hole binaries in hard X-ray states. Clearly, 4U 1636-536 became significantly harder during the low epoch, as the hard X-ray luminosity (L$_{X:20-80keV}\sim3.2\times10^{36}$\ erg\ s$^{-1}$) composed roughly half of the bolometric X-ray flux (2.0-80 keV, see table \ref{Spectral Data}).\\

\subsection{Comparison with X-ray transients}
The observed X-ray variability phenomena from 4U1636-536 resemble the main X-ray spectral states of black hole X-ray binaries, namely the ``hard'' and ``thermal'' states \citep{Done2003, Remillard2006}. Currently, various flavours of accretion flow models have been developed, mainly for black hole X-ray binaries, to not only explain these main spectral states, but also the more subtle states observed \citep[e.g. see the review by][]{Done2007}.  The most commonly used explanation assumes that the optically thick inner disc is replaced by a hot, optically thin, flow when black hole X-ray binaries change to the ``hard'' state. Although a similar scenario is used to explain the state transition in atoll sources, there is a fundamental difference with their black hole cousins: the presence of an observable solid surface at the neutron star. With its frequent state changes and abundant archival data from RXTE, Swift and INTEGRAL, 4U1636-536 could be an ideal target to study the behaviour of the accretion flow in neutron star binaries, and thereby adapt the models that are currently used for the BHXRBs.\\

\subsection{Optical/X-ray correlation}
Our three-month optical monitoring campaign on 4U1636-536 reveals a clear modulation which was obviously correlated with the X-ray states. Not surprisingly, this is consistent with the currently accepted mechanism for LMXB optical counterparts that they are dominated by X-ray reprocessing on the accretion disc and heated face of the donor star by the central X-ray source \citep[e.g.][]{vanParadijs1994, Shahbaz1998}. Thus the observed optical component largely follows variations in the intrinsic X-ray flux.\\

Our campaign also reveals that the increasing hard X-rays ($\gtrsim$ 10 keV) does not seem to influence the optical modulation. This is in contrast to LMC X-2 where its optical/X-ray correlation was much stronger with the hard X-rays than the soft \citep{McGowan2003}, suggesting that different accretion mechanisms are working in the inner regions of these systems.\\

One possible scenario that could explain our observation of 4U 1636-536 is Compton reflection \citep[see][]{Magdziarz1995, Nowak2003}. In this case, the softer part of X-ray continuum (power-law component with $\Gamma\sim$\ 2.12) are photoelectrically absorbed by material in the disc/stellar surface, while a large fraction of the hard X-rays are scattered out of the system with little change in photon energy. Since this would not explain the observations of LMC X-2, we think that further research into this topic is warranted, especially given the improvement in hard X-ray spectroscopic sensitivity over the last decade.


\section{Conclusions}
We have demonstrated that 4U1636-536 exhibits a coherent long-term variation across a wide range of wavelengths. The soft and hard X-rays are clearly anti-correlated on the $\sim$30-50d variability timescale, and this appears to be a persistent phenomenon. The evolution of the broad-band spectral characteristics of 4U 1636-536 resembles the {\it hard} and {\it soft/thermal} state change usually seen in black hole X-ray binaries.\\

Our work suggests that 4U 1636-536 is a unique and valuable target to understand the physics of X-rays binaries, including the size and location of the hard component in neutron star LMXBs, the interaction of hard X-rays with their surrounding environments such as the accretion disc/companion star surface, and the similarity or dissimilarity of accretion disc evolution between neutron star and black hole X-ray binaries. We are planning to explore these topics through a comprehensive optical spectroscopic and broad-band X-ray spectral analysis covering different variation epochs in the future.\\

\section{Acknowledgments}
We acknowledge the quick-look results provided by the ASM/RXTE team\footnote{including all those working on the ASM at MIT and at the Goddard Space Flight Center SOF and GOF.}, and the Swift/BAT transient monitor results provided by the Swift/BAT team\footnote{including all those working on the BAT at the Goddard Space Flight Centre and Los Alamos National Laboratory}. ICS would like to thank the Department of Physics and Astronomy, Michigan State University for providing the support and accessing the 1.3m telescope managed by SMARTS Consortium. RC acknowledges a Ramon y Cajal fellowship (RYC-2007-01046) and a Marie Curie European Reintegration Grant (PERG04-GA-2008-239142).  We would also like to thank for the anonymous reviewer's valuable comments.\\

\end{document}